\documentclass{article}

\newtheorem{definition}{Definition}[section]

\usepackage{array, latexsym, graphicx,times,amsfonts,picture,color,multirow,epsf,epsfig,mathtools}
\usepackage{amsmath}
\usepackage{verbatim}
\usepackage{blkarray}
\usepackage{setspace}
\usepackage{subfigure}
\usepackage{breqn}
\usepackage{listings}
\usepackage{enumitem}
\usepackage[T1]{fontenc}
\usepackage{textcomp}
\usepackage{hyperref}

\usepackage[utf8]{inputenc}

\begin{document}
\newtheorem{example}{Example}
\title{Automated Market Makers for Decentralized Finance (DeFi)}
\author{Yongge Wang\\ UNC Charlotte}

\maketitle

\begin{abstract}
This paper compares mathematical models for automated market makers (AMM) including
logarithmic market scoring rule (LMSR), liquidity sensitive LMSR (LS-LMSR),
constant product/mean/sum, and others. It is shown that though LMSR may not be a good model
for Decentralized Finance (DeFi) applications, LS-LMSR has several advantages over 
constant product/mean based AMMs. 
This paper proposes and analyzes constant ellipse based cost functions for 
AMMs. The proposed cost functions are computationally
efficient (only requires multiplication and square root calculation) and have certain advantages
over widely deployed constant product cost functions. For example, the proposed
market makers are more robust against slippage based front running attacks. In addition 
to the theoretical advantages of constant ellipse based cost functions,
our implementation shows that if the model is used as a cryptographic property
swap tool over Ethereum blockchain, it saves up to 46.88\% gas cost against 
Uniswap V2 and saves up to 184.29\% gas cost against Uniswap V3 which has been launched in 
April 2021.  The source codes related to this paper are available at 
\url{https://github.com/coinswapapp}
and the prototype of the proposed AMM is available at
\url{https://yonggewang.github.io/ethcoinswap/}.
\end{abstract}

\section{Introduction}
Decentralized finance (DeFi or open finance) is implemented through smart contracts (DApps)
which are stored on a public distributed ledger (such as a blockchain)
and can be activated to automate execution of financial instruments and digital assets. The immutable 
property of blockchains guarantees that these DApps are also tamper-proof and the content 
could be publicly audited.

DeFi applications range from automated markets (e.g., Uniswap \cite{uniswapprotocol} and Curve Finance), 
price oracles (e.g., Chainlink), to financial derivatives and many others.
Most DeFi applications (e.g., Bancor \cite{hertzog2017bancor} 
and  Compound \cite{compoundpprotocol})
enable smart token transaction instantly by using price equilibrium mechanisms based on total availability supply
(or called bonding curves), though still some of DeFi applications do not carry out instant transaction.
In a blockchain system, traders submit 
their transactions to the entire blockchain network (e.g., stored in the mempool), 
a miner in the system collects these transactions, 
validates them, and puts them into  a valid block that is eventually added to an immutable chain of blocks. 
These submitted transactions (e.g., the mempool for Ethereum could be viewed at 
\url{https://etherscan.io/txsPending}) are visible to all nodes. A malicious node (the miner itself could be malicious) 
may construct his/her own malicious transactions
based on these observed transactions and insert her malicious transactions 
before or after the observed transactions by including 
appropriate gas costs (see, e.g., \cite{journals/corr/abs-2009-14021}). 
These malicious transactions take Miner Extractable Value (MEV) profit with minimal cost.
With their own experience of failing to recover some tokens of 12K USD value
in a Uniswap V2 pair (these tokens were recovered by a front running bot), 
Robinson and  Konstantopoulos \cite{darkforesteth} describe the Ethereum blockchain
as a Dark Forest. The flashbots website (\url{https://explore.flashbots.net/})
shows that the total extracted MEV by front running bots in the 24 hours of May 18, 2021 is around 
8.6M USD. In addition to the front running attacks, it is also common to mount attacks against DeFi price oracles.
In the DeFi market, a lender (a smart contract) normally queries an oracle to determine the fair market value (FMV) 
of  the borrower's collateral. 

This paper analyzes existing mathematical models for AMMs and discusses 
their applicability to blockchain based DeFi applications.
One important consideration for the discussion is to compare the model resistance to front running attacks.  
Our analysis shows that though LS-LMSR is the best among existing models, 
it may not fit the blockchain DeFi application due to the following two reasons:
\begin{itemize}
\item LS-LMSR involves complicated computation and it is not gas-efficient for DeFi implementations.  
\item The cost function curve for LS-LMSR market is concave.  In order to reflect the DeFi market 
principle of supply and demand, it is expected that the cost function 
curve should be convex. 
\end{itemize}
Constant product based Uniswap AMM has been very successful as a DeFi swapping application.  
However,  our analysis shows that Uniswap V2 \cite{uniswapprotocol} has a high slippage 
(in particular, at the two ends) and may not be a best choice for several applications. 
This paper proposes a constant ellipse based
AMM model.  It achieves the same model property as LS-LMSR but its cost function curve is 
convex and it is significantly gas-efficient for DeFi applications. 
At the same time, it reduces the sharp price flunctuation challenges by Uniswap V2.
We have implemented and deployed a prototype CoinSwap based on our constant ellipse AMM 
during March 2021 (see \url{https://yonggewang.github.io/ethcoinswap/}) and released a technical report \cite{wang2020automated} of this paper 
during September 2020.
The CoinSwap has a controllable slippage mechanism and has a mechanisms for Initial Coin Offer (ICO).
It should be noted that Uniswap team was aware of their Uniswap V2 disadvantages that we have just mentioned and,
independent of this paper, 
proposed the Uniswap V3 \cite{uniswapprotocolv3} (released during April 2021). 
Uniswap V3 tried to address this challenge by using a  shifted constant product equation $(x+\alpha)(y+\beta)=K.$
Though this shifted equation in Uniswap V3 resolves some of the challenges that constant ellipse 
AMM has addressed and it can implement some of the funcationalities in CoinSwap (e.g., 
ICO and reduced slippage),  it still does not have  a smooth price flunctuation at two ends. 
Furthermore,  the experimental data from CoinSwap project
shows that Uniswap V3 has significant high gas costs than CoinSwap (to achieve the same functionality). 

The structure of the paper is as follows. Section \ref{reviewsec} gives an introduction to 
prediction markets and analyzes various models for automated market makers (AMM).
Section \ref{coinswapsec} proposes a new constant ellipse market maker model.
Section \ref{pricecomsec} compares various cost functions from aspects of the principle of supply and demand, 
coin liquidity, and token price fluctuation. 
Section \ref{lslmsrexsec} compares price amplitude for various cost functions and 
Section \ref{impelsec} discusses the implementation details. 

\section{Existing models for prediction market makers}
\label{reviewsec}
\subsection{Prediction market and market makers}
It is commonly believed that combined information/knowledge of all traders 
are incorporated into stock prices immediately (Fama \cite{fama1970efficient} includes 
this as one of his ``efficient market hypotheses'') . For example, these information may be used by traders
to hedge risks in financial markets such as stock and commodities future markets.
With aggregated information from all sources, speculators who seek to ``buy low and sell high'' can 
take profit by predicting future prices from current prices and aggregated information.
Inspired by these research,  the concept of ``information market'' was introduced to investigate the common
principles in information aggregation. Among various approaches to information market, 
a {\em prediction market} is an exchange-traded market for the purpose of eliciting aggregating beliefs 
over an unknown future outcome of a given event. As an example, in a horse race with $n$ horses,
one may purchase a security of the form ``horse $A$ beats horse $B$''. This security pays off \$1 if 
horse $A$ beats horse $B$ and \$0 otherwise. Alternatively, one may purchase other securities 
such as ``horse $A$ finishes at a position in $S$'' where $S$ is a subset of $\{1, \cdots, n\}$.
For the horse race event, the outcome space consists of the $n!$ possible permutations
of the $n$ horses.

For prediction markets with a huge outcome space, the continuous double-sided auction  
(where the market maker keeps an order book that tracks bids and asks) may fall victim of the 
{\em thin-market} problem. Firstly, in order to trade, traders need to coordinate on what or when 
they will trade. If there are significantly less participants than the size of the outcome space,
the traders may only expect substantial trading activities in a small set of assets and many assets 
could not find trades at all. Thus the market 
has a low to poor liquidity. Secondly, if a single participant knows something about an event
while others know nothing about this information, this person may choose not to release this 
information at all or only release this information gradually. This could be justified as follows.
If any release of this information (e.g., a trade based on this information) is a signal to other 
participants that results in belief revision discouraging trade, the person may choose not to release 
the information (e.g., not to make the trade at all). On the other hand, this person
may also choose to leak the information into the market gradually over time to obtain a greater profit.
The second challenge for the standard information market is due to the {\em irrational participation}
problem where a rational participant may choose not to make any speculative trades with 
others (thus not to reveal his private information) after hedging 
his risks derived from his private information. 

\subsection{Logarithmic market scoring rules (LMSR)}
{\em Market scoring rules}
are commonly used to overcome the thin market and the irrational participation problems
discussed in the preceding section. Market scoring rule based automated market makers (AMM)
implicitly/explicitly maintain prices for all assets at certain prices and are willing to trade on every assets.
In recent years, Hanson's logarithmic market scoring rules (LMSR) AMM 
\cite{hanson2003combinatorial,hanson2007logarithmic}
has become the de facto AMM mechanisms for prediction markets.

Let $X$ be a random variable with a finite outcome space $\Omega$.
Let ${\bf p}$ be a reported probability estimate for the random variable $X$.
That is, $\sum_{\omega\in\Omega}{\bf p}(\omega)=1$. 
In order to study rational behavior (decision) with fair fees, Good \cite{good1992rational}
defined a reward function with the logarithmic market scoring rule (LMSR) as follows:
\begin{equation}
\label{refunctioneq}
\left\{s_\omega({\bf p})=b\ln (2\cdot {\bf p}(\omega))\right\}
\end{equation}
where $b>0$ is a constant.  A participant in the market may choose to change the current 
probability estimate ${\bf p}_1$ to a new estimate ${\bf p}_2$. This participant will be 
rewarded $s_\omega({\bf p}_2)-s_\omega({\bf p}_1)$ if the outcome $\omega$ happens.
Thus the participant would like to maximize his expected value (profit)
\begin{equation}
\label{inentropy}
S({\bf p}_1, {\bf p}_2)=\sum_{\omega\in\Omega}{\bf p}_2(\omega)\left(s_\omega({\bf p}_2)-s_\omega({\bf p}_1)\right)
=b\sum_{\omega\in\Omega}{\bf p}_2(\omega)\ln\frac{{\bf p}_2(\omega)}{{\bf p}_1(\omega)}
=bD({\bf p}_2||{\bf p}_1)
\end{equation}
by honestly reporting his believed probability estimate, where $D({\bf p}_2||{\bf p}_1)$ is the relative entropy or 
Kullback Leibler distance between the two probabilities ${\bf p}_2$ and ${\bf p}_1$.
An LMSR market can be considered as a sequence of logarithmic scoring rules
where the market maker (that is, the patron) pays the last participant and receives payment from the first participant.

Equivalently, an LMSR market can be interpreted as a market maker offering $|\Omega|$ securities
where each security corresponds to an outcome and pays \$1 if the outcome 
is realized \cite{hanson2003combinatorial}. In particular, changing the market probability 
of $\omega\in\Omega$ to a value ${\bf p}(\omega)$ is equivalent to buying the security for $\omega$
until the market price of the security reaches ${\bf p}(\omega)$. As an example for the decentralized financial (DeFi)
AMM on blockchains, assume that the market maker offers $n$ categories of tokens.
Let ${\bf q}=(q_1, \cdots, q_n)$ where $q_i$ represents the number of outstanding tokens 
for the token category $i$.
The market maker keeps track of the cost function $C({\bf q})=b\ln\sum_{i=1}^ne^{q_i/b}$
and a price function for each token
\begin{equation}
\label{pricefunc}
P_i({\bf q})=\frac{\partial C({\bf q})}{\partial q_i}= \frac{e^{q_i/b}}{\sum_{j=1}^ne^{q_j/b}}
\end{equation}
It should be noted that the equation (\ref{pricefunc}) is a generalized inverse of the scoring rule function 
(\ref{refunctioneq}).
The cost function captures the amount of total assets wagered in the market where 
$C({\bf q}_0)$ is the market maker's maximum subsidy to the market. The price function
$P_i({\bf q})$ gives the current cost of buying an infinitely small quantity of the 
category $i$ token. If a trader wants to change the number of outstanding shares from 
${\bf q}_1$ to ${\bf q}_2$, the trader needs to pay the cost difference $C({\bf q}_2)-C({\bf q}_1)$.


Next we use an example to show how to design AMMs using LMSR.
Assume that $b=1$ and the patron sets up an automated market marker ${\bf q}_0=(1000,1000)$
by depositing 1000 coins of token $A$ and 1000 coins of token $B$. The initial market cost is 
$C({\bf q}_0)=\ln \left(e^{1000}+e^{1000}\right)=1000.693147.$
The instantaneous prices for a coin of tokens are
$P_A({\bf q}_0)=\frac{e^{1000}}{e^{1000}+e^{1000}}=0.5$ and 
$P_B({\bf q}_0)=\frac{e^{1000}}{e^{1000}+e^{1000}}=0.5$.
If this AMM is used as a price oracle, then one coin of token $A$ equals 
$\frac{P_A({\bf q}_0)}{P_B({\bf q}_0)}=1$ coin of token $B$. 
If a trader uses 0.689772 coins of token $B$ to buy 5 coins of token $A$ from market ${\bf q}_0$,
then the market  moves to a state ${\bf q}_1=(995,1000.689772)$ with a 
total market cost $C({\bf q}_1)=1000.693147=C({\bf q}_0)$. The instantaneous prices for a coin of tokens 
in ${\bf q}_1$ are
$P_A({\bf q}_1)=0.003368975243$ and
$P_B({\bf q}_1)=295.8261646$.
Now a trader can use 0.0033698 coins of token $B$ to purchase 995 coins of token
$A$ from the AMM ${\bf q}_1$ with a resulting market maker state
${\bf q}_2=(0,1000.693147)$ and a total market cost $C({\bf q}_2)=1000.693147=C({\bf q}_0)$.

\begin{figure}[htb]
\caption{LMSR market maker cost function curves for $C(x,y,z)=100$ and $C(x,y)=100$}
\label{lmsr3dfig}
\begin{center}
\includegraphics[width=0.45\textwidth]{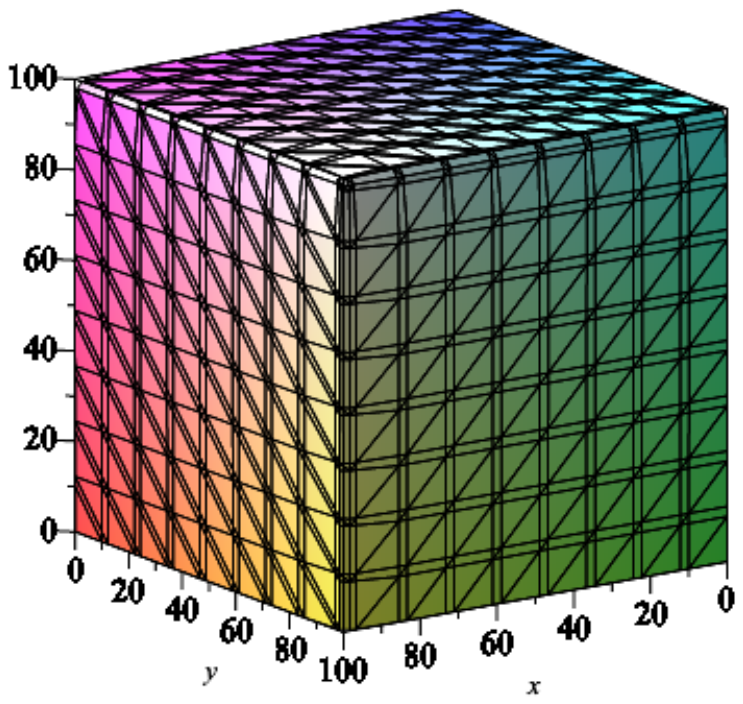} \!\!\!\!\!\!\!\!\!\!\!\!\!\!\!\!\!\!\!\!\!\
\includegraphics[width=0.45\textwidth]{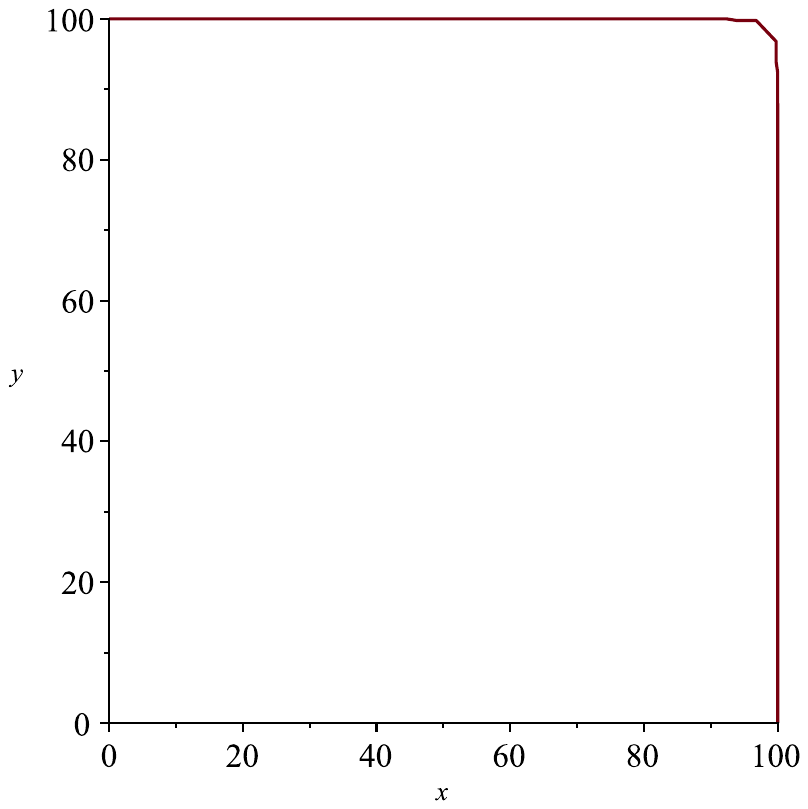} 
\end{center}
\end{figure}

The above example shows that LMSR based AMM 
works well only when the outstanding shares of the tokens are evenly distributed 
(that is, close to 50/50). 
When the outstanding shares of the tokens are not evenly distributed,  a trader can 
purchase all coins of the token with lesser outstanding shares
and let the price ratio $\frac{P_A({\bf q})}{P_B({\bf q})}$ change to an arbitrary 
value with a negligible cost. This observation is further
justified by the LMSR cost function curves in Figure \ref{lmsr3dfig}.
The first plot is for the cost function $C(x,y,z)=100$ with three tokens and 
the second plot is for the cost function $C(x,y)=100$ with two tokens. The second plot 
shows that the price for each token fluctuates smoothly only
in a tiny part (the upper-right corner) of the curve with evenly distributed token shares. Outside of this part,
the tangent line becomes vertical or horizontal. That is, one can use a tiny amount of one token
to purchase all outstanding coins of the other token in the market maker. In a conclusion,
LMSR based AMMs may not be a good solution for DeFi applications.

\vskip -3cm

\noindent
In the traditional prediction market, the three desired properties for a pricing rule to have include:
{\em path independence, translation invariance,} and {\em liquidity sensitivity}.
Path independence means that if the market moves from one state to another state, the payment/cost is
independent of the paths that it moves.  If path independence is not achieved,  the adversary trader may 
place a series of transactions along a calculated path and obtain profit without any risk.  Thus this is an essential
property that needs to be satisfied.  An AMM with a cost function generally achieves path independence.
Thus all models that we will analyze in this paper (including our proposed constant ellipse AMM model) achieve 
path independence. On the other hand, the translation invariance guarantees that no trader can arbitrage
the market maker without risk by taking on a guaranteed payout for less than the payout.
As an example,,  a translation invariant pricing rule preserves the equality between the price of an event 
and the probability of that event occurring.  Translation invariant rules also guarantee the ``law of one price'' 
which says that if two bets offer the same payouts in all states, they will have the same price. 
Liquid sensitivity property requries that a market maker should adjust the elasticity of 
their pricing response based on the volume of activity in the market.  For example,
as a generally marketing practice,  this property requires that a fixed-size investment moves prices 
less in thick (liquid) markets than in thin (illiquid) markets. Though liquid sensitivity is a nice property 
to be achieved, a healthy market maker should not be too liquid sensitive (in our later examples,
we show that Uniswap V2 is TOO liquid sentitive).

\begin{figure}[htb]
\caption{LS-LMSR market maker cost function curves for $C(x,y,z)=100$ and $C(x,y)=100$}
\label{lslmsr3dfig}
\begin{center}
\includegraphics[width=0.43\textwidth]{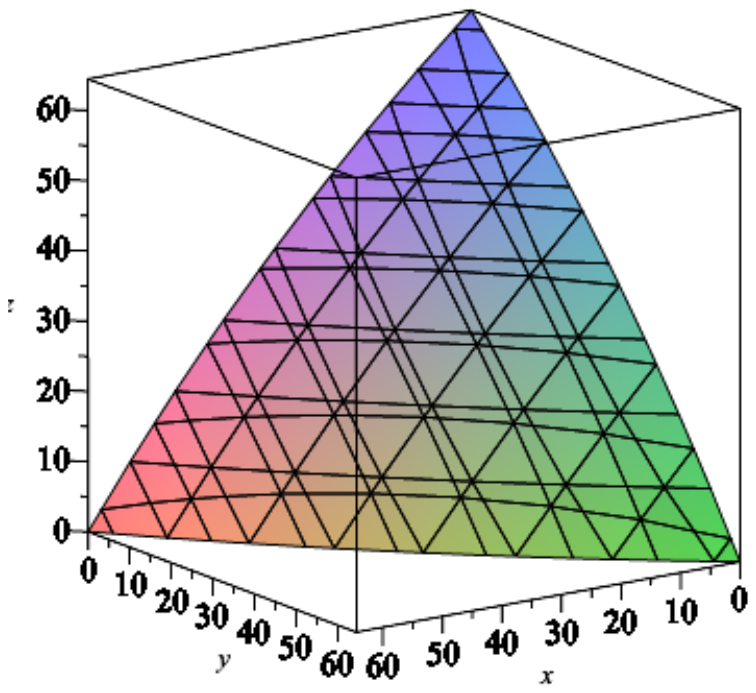} \!\!\!\!\!\!\!\!\!\!\!\!\!\!\!\!\!\!\!\!\!\!\!\
\includegraphics[width=0.43\textwidth]{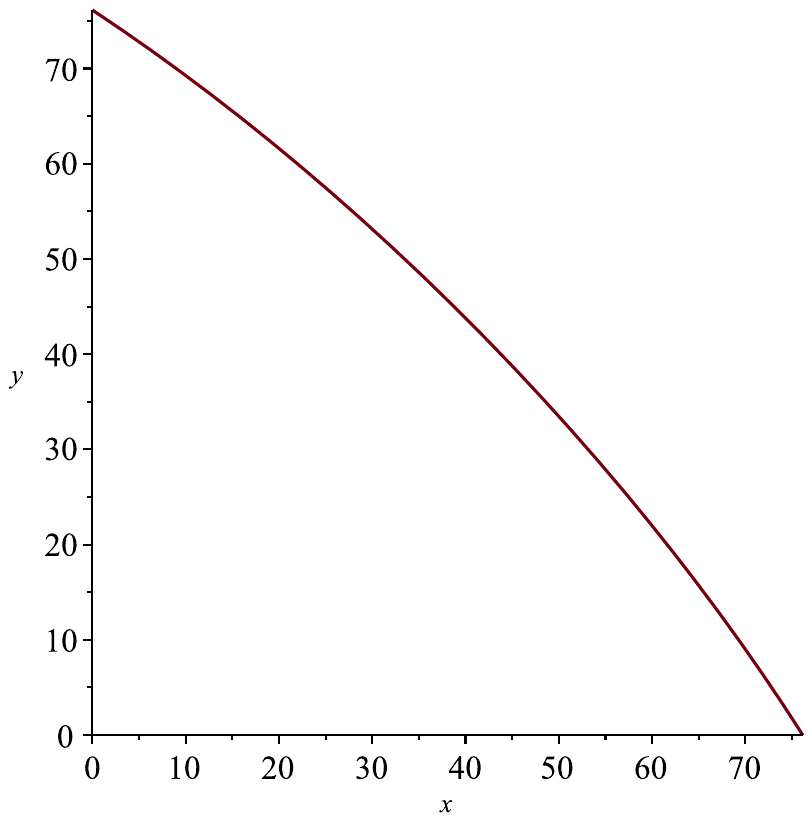} 
\end{center}
\end{figure}

\begin{definition}
(see, e.g., Othman et al \cite{othman2013practical}) For a pricing rule $P$,
\begin{enumerate}
\item $P$ is path independent if the value of line integral (cost) between any two quantity vectors depends only on those quantity vectors, and not on the path between them.
\item $P$ is translation invariant if  
$\sum_i P_i({\bf q})=1$
for all valid market state ${\bf q}$.
\item $P$  is liquidity insensitive if
$P_i({\bf q}+(\alpha, \cdots, \alpha))=P_i({\bf q})$
for all valid market state ${\bf q}$ and $\alpha$. $P$  is liquidity sensitive if it is not  liquidity insensitive.
\end{enumerate}
\end{definition}
Othman et al \cite{othman2013practical} showed that no market maker can satisfy all three of the desired properties
at the same time. Furthermore, Othman et al \cite{othman2013practical} showed that LMSR satisfies translation
invariance and path independence though not liquidity sensitivity.  In practice, the patron would prefer liquidity sensitity instead
of absolute translation invariance. By relaxing the translation invariance to $\sum_i P_i({\bf q})\ge 1$,
Othman et al \cite{othman2013practical} proposed the Liquidity-Sensitive LMSR market. In particular,
LS-LMSR changes the constant $b$ in the LMSR formulas
to $b({\bf q})=\alpha\sum_i q_i$ where $\alpha$ is a constant and requiring the cost function to always
move forward in obligation space.  Specifically, for ${\bf q}=(q_1, \cdots, q_n)$, the market maker 
keeps track of the cost function $C({\bf q})=b({\bf q})\ln\sum_{i=1}^ne^{q_i/b({\bf q})}$
and a price function for each token
\begin{equation}
\label{lspricefunc}
P_i({\bf q})=\alpha\ln\left(\sum_{j=1}^ne^{q_j/b({\bf q})}\right)+ 
\frac{e^{q_i/b({\bf q})}\sum_{j=1}^nq_j - \sum_{j=1}^nq_j e^{q_j/b({\bf q})}  }{\sum_{j=1}^nq_j \sum_{j=1}^n e^{q_j/b({\bf q})}}
\end{equation}
Furthermore, in order to always move forward in obligation space, we need to revise the cost 
that a trader should pay. In the proposed ``no selling'' approach, assume that the market is at state ${\bf q}_1$ 
and the trader tries to impose an obligation ${\bf q}_\delta=(q'_{1}, \cdots, q'_{n})$ 
to the market with $\bar{q}_\delta=\min_iq'_{i}<0$.
That is, the trader puts $q'_i$ coins of token $i$ to the market if $q'_i\ge 0$ and 
receives $-q'_i$ coins of token $i$ from the market if $q'_i< 0$. 
Let $\bar{\bf q}_\delta=(-\bar{q}_\delta, \cdots, -\bar{q}_\delta)$. Then the trader should pay 
$C({\bf q}+{\bf q}_\delta+\bar{\bf q}_\delta)+\bar{q}_\delta-C({\bf q})$
and the market moves to the new state ${\bf q}+{\bf q}_\delta+\bar{\bf q}_\delta$.
In the proposed ``covered short selling approach'', the market moves in the same way as LMSR market except
that if the resulting market ${\bf q}'$ contains a negative component, then the market ${\bf q}'$ automatically 
adds a constant vector to itself so that all components are non-negative. 
In either of the above proposed approach, if ${\bf q}+{\bf q}_\delta$ contains negative components, 
extra shares are automatically mined and added to the market to avoid negative outstanding shares.
This should be avoided in DeFi applications. In DeFi applications,
one should require that ${\bf q}_\delta$ could be imposed to a market  ${\bf q}_0$ only if there is no 
negative component in ${\bf q}+{\bf q}_\delta$ and the resulting market state is ${\bf q}+{\bf q}_\delta$. 
LS-LMSR is obviously path independent since it has a cost function.
Othman et al \cite{othman2013practical} showed that LS-LMSR has the desired liquidity sensitive property.
On the other hand, LS-LMSR satisfies the relaxed translation invariance $\sum_i P_i({\bf q})\ge 1$.  This means that 
if a trader imposes an obligation and then sells it back to the market maker, the trader may end up with a net lost 
(this is similar to the  markets we see in the real world).
Figure \ref{lslmsr3dfig} displays the curve of the cost function $C(x,y,z)=100$ for LS-LMSR market maker 
with three tokens and the curve of the cost function $C(x,y)=100$ for LS-LMSR market maker 
with two tokens. It is clear that these two curves are concave.

\vskip -3cm

\subsection{Constant product/sum/mean AMMs}
\begin{figure}[htb]
\caption{Constant product cost function curves for $xyz=100$ and $xy=100$}
\label{constant3dfig}
\begin{center}
\includegraphics[width=0.45\textwidth]{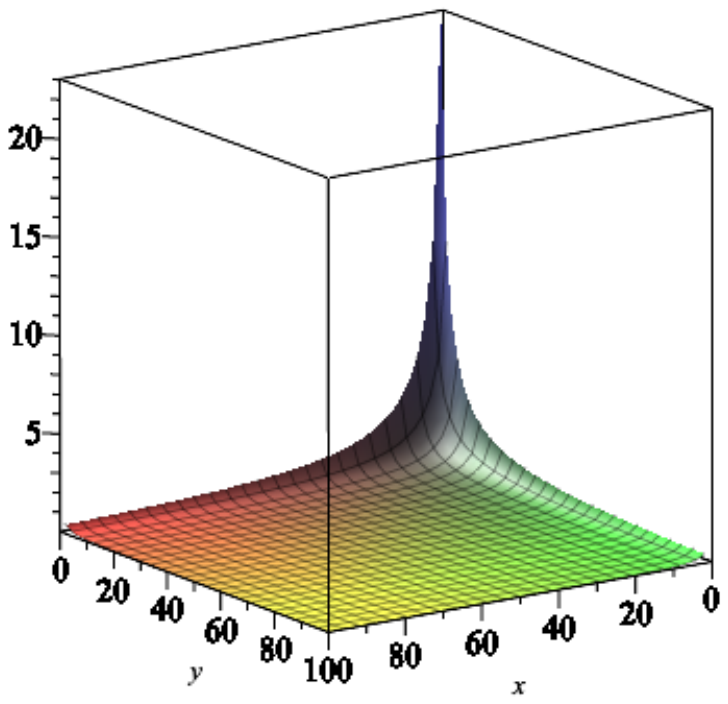} \!\!\!\!\!\!\!\!\!\!\!\!\!\!\!\!\!\!\!\!\!\!
\includegraphics[width=0.45\textwidth]{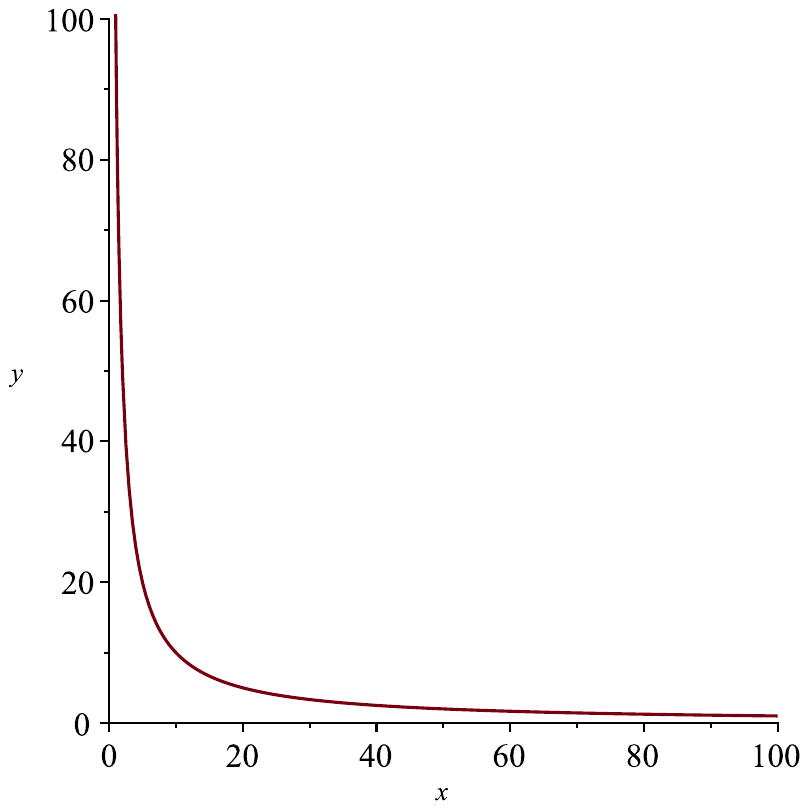} 
\end{center}
\end{figure}

Constant product market makers have been used in DeFi applications (e.g., Uniswap \cite{uniswapprotocol})  
to enable on-chain exchanges of digital assets and
on-chain-decentralized price oracles. 
In this market, one keeps track of the cost function
$C({\bf q})={\prod_{i=1}^nq_i}$ as a constant. For this market, the price function for each token
is defined as
$P_i({\bf q})=\frac{\partial C({\bf q})}{\partial q_i}={\prod_{j\not= i}q_j}.$
Figure \ref{constant3dfig} shows the curve of the constant product cost function $xyz=100$
with three tokens and the curve of the constant product cost function $xy=100$ with two tokens.

\vskip -3cm

\begin{figure}[htb]
\caption{Constant mean cost function curves for $xy^2z^3=100$ and $x^2y^3=100$}
\label{constantmean3dfig}
\begin{center}
\includegraphics[width=0.45\textwidth]{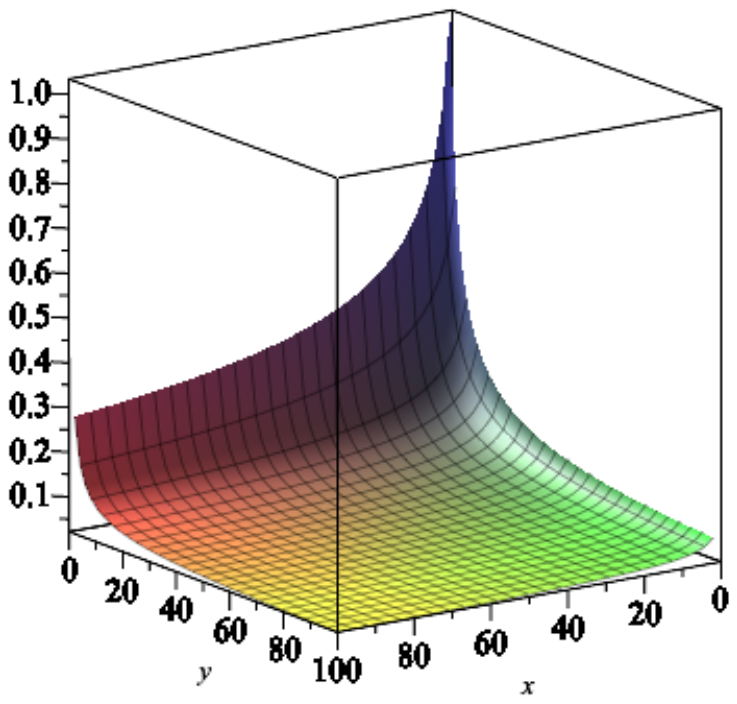}  \!\!\!\!\!\!\!\!\!\!\!\!\!\!\!\!\!\!\!\!\!\!\!\!\!\!\!\!\
\includegraphics[width=0.45\textwidth]{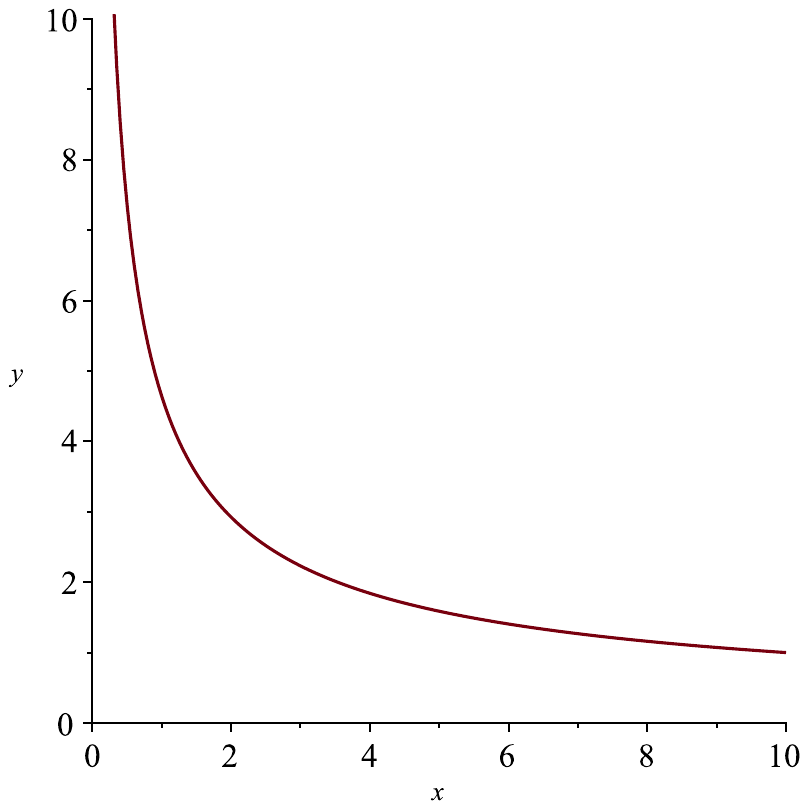} 
\end{center}
\vskip -3cm
\end{figure}

The cost function $C({\bf q})={\prod_{i=1}^nq_i^{w_i}}$
has been used to design constant mean AMMs \cite{balancer} 
where $w_i$ are positive real numbers.  In the constant mean market,
the price function for each token is 
$P_i({\bf q})=\frac{\partial C({\bf q})}{\partial q_i}={w_iq_i^{w_i-1}\prod_{j\not= i}q_j}.$
Figure \ref{constantmean3dfig} shows the curve of the constant mean cost function $xy^2z^3=100$
with three tokens and the curve of the constant mean cost function $x^2y^3=100$ with two tokens.

\begin{figure}[htb]
\caption{Constant sum market maker cost function curves for $x+y+z=100$ and $x+y=100$}
\label{constants3dfig}
\begin{center}
\includegraphics[width=0.45\textwidth]{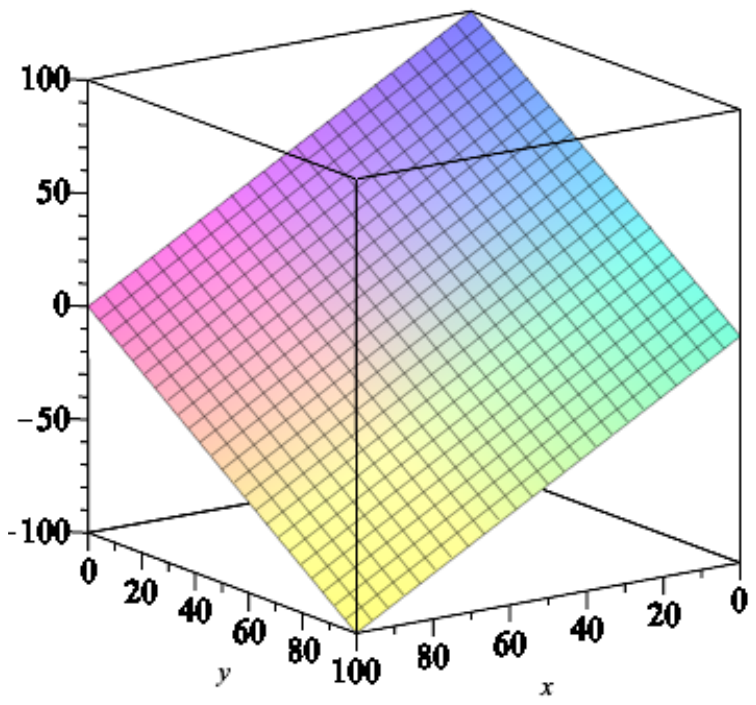}  \!\!\!\!\!\!\!\!\!\!\!\!\!\!\!\!\!\!\!\!\!\!\!\!\!\!\!\!\!\!\!\!\!
\includegraphics[width=0.45\textwidth]{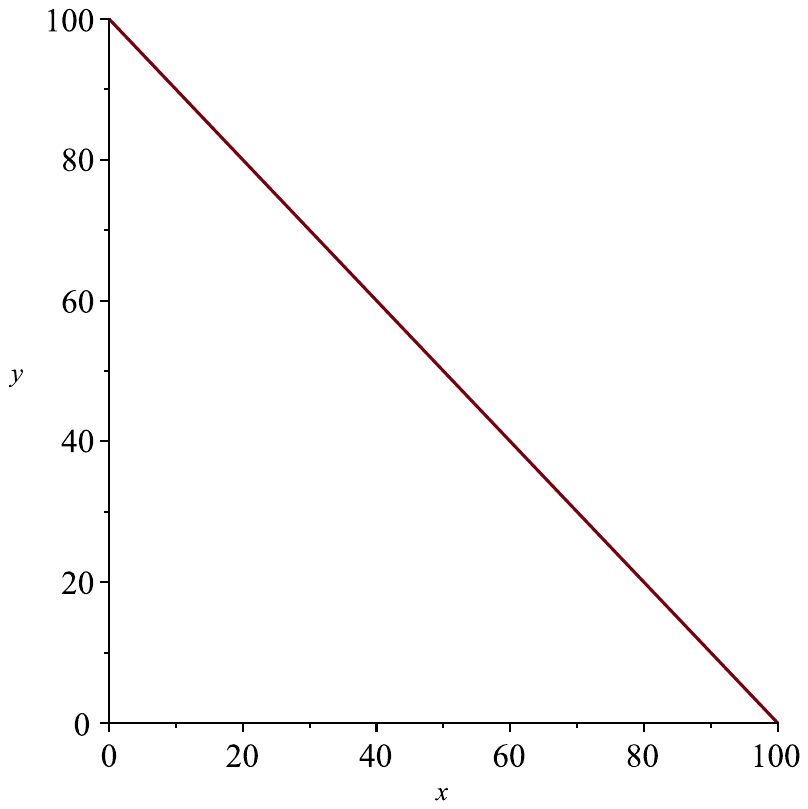} 
\end{center}
\vskip -3cm
\end{figure}

One may also use the cost function $C({\bf q})={\sum_{i=1}^nq_i}$ to design 
constant sum market makers. In this market, the price for each token is always $1$. 
That is, one coin of a given token can be used to trade for 
one coin of another token at any time when supply lasts.  
Figure \ref{constants3dfig} shows the 
curve of the constant sum cost function $x+y+z=100$
with three tokens and the curve of the constant sum cost function $x+y=100$ with two tokens.

The curves in Figures \ref{constant3dfig} and \ref{constantmean3dfig} show that 
constant product/mean/sum AMMs are highly liquidity sensitive when the distribution 
of the tokens are far from balanced market states (where the price flunctuates sharply).
By the fact that there exist cost functions,  constant product/mean/sum AMMs achieve
path independence.  It is also straightforward to check that constant product/mean AMMs 
are liquidity sensitive.  By the fact (see \cite{othman2013practical}) that no market maker can satisfy all three of 
the desired properties at the same time, constant product/mean AMMs  are not translation invariant. 
It is also straightforward to check that the constant sum AMM is liquidity insensitive.  Since  liquidity sensitivity
is one of the essential market rules to be satisfied, in the remaining part of this paper, we will no long discuss
constant sum models.

\section{Constant ellipse AMMs}
\label{coinswapsec}
\begin{figure}[htb]
\caption{Constant ellipse cost function curves for three and two tokens}
\label{constantcircle3dfig}
\begin{center}
\includegraphics[width=0.45\textwidth]{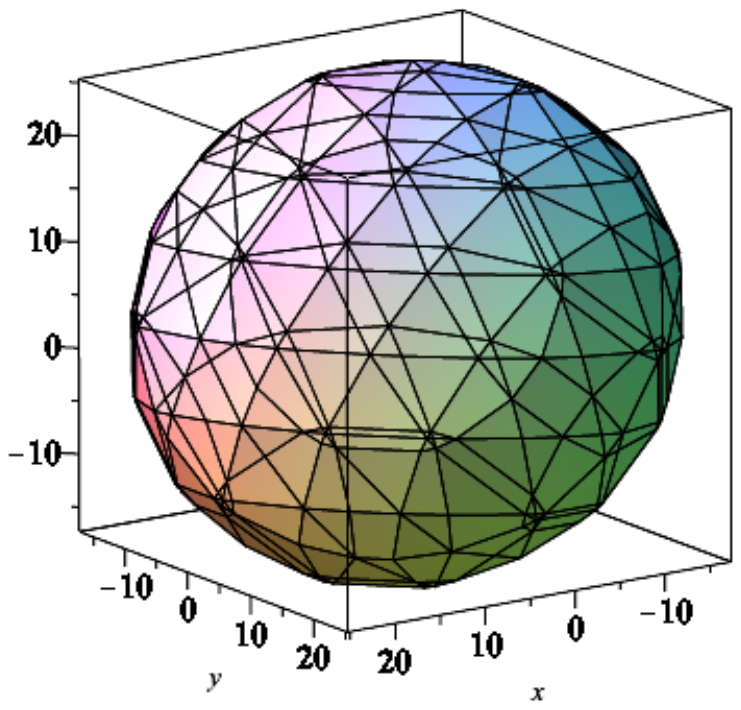} \!\!\!\!\!\!\!\!\!\!\!\!\!\!\!\!\!\!\!
\includegraphics[width=0.45\textwidth]{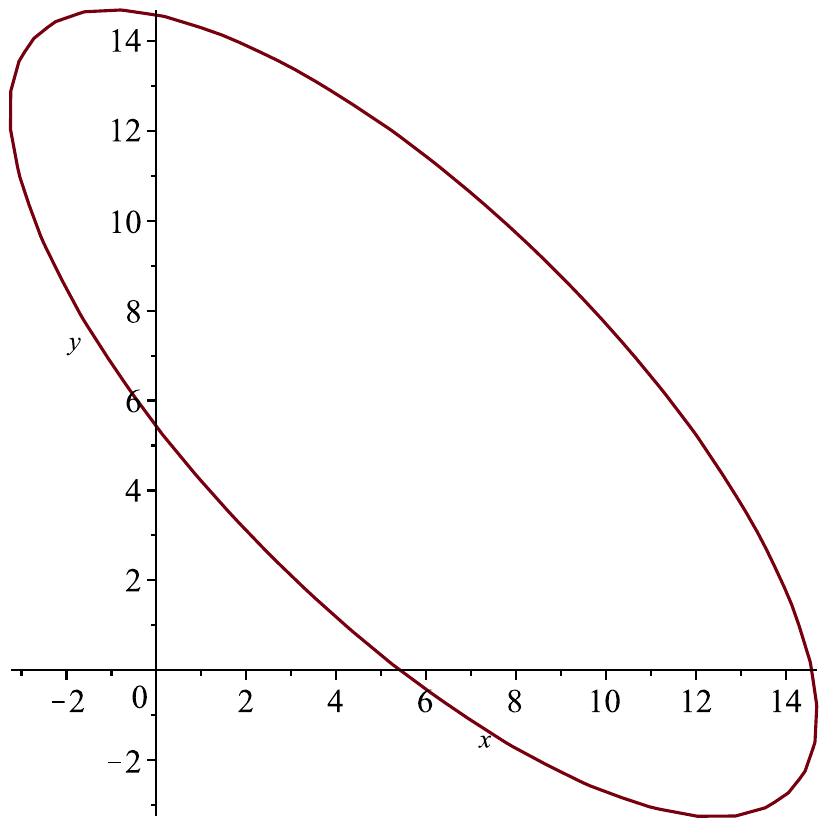} 
\end{center}
\vskip -3cm
\end{figure}

Section \ref{pricecomsec} compares the advantages and disadvantages of LMSR, LS-LMSR, 
and constant product/mean/sum AMMs. The analysis shows that none of them is ideal for 
DeFi applications. In this section, we propose AMMs based on constant 
ellipse cost functions. That is, the AMM's cost function is defined by 
\begin{equation}
\label{constantcircleeq}
C({\bf q})={\sum_{i=1}^n (q_i-a)^2+b\sum_{i\not=j}q_iq_j}
\end{equation}
where $a, b$ are constants. The price function for each token is 
$$P_i({\bf q})=\frac{\partial C({\bf q})}{\partial q_i}={2(q_i-a)+b\sum_{j\not=i}q_j}.$$
For AMMs, we only use the first quadrant of the coordinate plane.
By adjusting the parameters $a,b$ in the equation (\ref{constantcircleeq}),
one may keep the cost function to be concave (that is, using the upper-left part of the ellipse)
or to be convex (that is, using the lower-left part of the ellipse). By adjusting the absolute value 
of $a$, one may obtain various price amplitude and price fluctuation rates based on the principle of supply and demand
for tokens. It is observed that constant ellipse AMM price functions are liquidity sensitive
and path independent but not translation invariance.
Figure \ref{constantcircle3dfig} shows the curve of the constant ellipse cost function 
$$(x-10)^2+(y-10)^2+(z-10)^2+1.5(xy+xz+yz)=350$$
with three tokens and the curve of the the constant ellipse cost function 
$$(x-10)^2+(y-10)^2+1.5xy=121$$
with two tokens. As mentioned in the preceding paragraphs, one may use convex or concave part of the 
ellipse for the cost function. For example, in the second plot of Figure \ref{constantcircle3dfig}, 
one may use the lower-left part in the first quadrant as a convex cost function 
or use the upper-right part in the first quadrant as a concave cost function.
It is straightforward to verity that the constant ellipse AMM achieve
path independence and liquidity sensitivity.  Though constant ellipse AMM
is not translation invariant,  our analysis and examples provide evidence that in a
constant ellipse AMM,  a trader have certain risks for arbitraging
the market maker on a payout for less than the payout (this is related to our analysis
on the slippage in the later sections).

\noindent

\section{Supply-and-demand, liquid sensitivity,  and price fluctuation}
\label{pricecomsec}
Without loss of generality, this section considers AMMs consisting of two tokens:
a USDT token where each USDT coin costs one US dollar and an imagined spade suit token $\spadesuit$.
The current market price of a $\spadesuit$ token coin could have different values
such as half a USDT coin, one USDT coin, two USDT coins, or others. 
In Decentralized Finance (DeFi) applications, the patron needs to provide liquidity by depositing
coins of both tokens in the AMM. 
Without loss of generality, we assume that, at the time when the AMM is incorporated, the market price
for a coin of spade suit token is equivalent to one USDT coin. 
For general cases that the market price for one $\spadesuit$ coin is not equivalent to one USDT coin
at the time when the market maker is incorporated, we can create virtual shares in the AMM
by dividing or merging actual coins. That is, each share of USDT (respectively $\spadesuit$) in the AMM 
consists of a multiple or a portion of  USDT (respectively $\spadesuit$) coins. 
One may find some examples in Section \ref{lslmsrexsec}. 

To simplify our notations, we will use ${\bf q}=(x,y)$ instead of ${\bf q}=(q_1,q_2)$
to represent the market state. In this section, we will only study 
the price fluctuation of the first token based on the principle of supply and demand 
and the trend of the price ratio $\frac{P_x({\bf q})}{P_y({\bf q})}$ which is strongly related liquid sensitivity. 
By symmetry of the cost functions, the price fluctuation of the second token 
and the ratio $\frac{P_y({\bf q})}{P_x({\bf q})}$ have the same property. 
In the following, we analyze the token price fluctuation for various AMM 
models with the initial market state ${\bf q}_0=(1000, 1000)$. That is, the patron creates 
the AMM by depositing 1000 USDT coins and 1000 spade suit coins in the market.
The analysis results are summarized in Table \ref{pricetabcom}.

\begin{table}[h!]
  \begin{center}
    \caption{Token price comparison}
    \label{pricetabcom}
    \begin{tabular}{@{}|c|l|l|l|l|@{}} \hline    
     \textbf{AMM type}&\textbf{market cost} & \textbf{${P_x({\bf q})}/{P_y({\bf q})}$}&\textbf{tangent $\frac{\partial y}{\partial x}$}\\ \hline
LS-LMSR &2386.29436&$(0.648, 1.543)$&(-1.543,-0.648) \\ \hline
cons. product &1000000&$(0,\infty)$&$(-\infty,0)$\\ \hline
cons. sum &2000&1& -1\\ \hline
cons. ellipse &50000000&$(0.624,1.604)$& $(-1.604, -0.624)$ \\ \hline
    \end{tabular}
  \end{center}
\end{table}

\subsection{LS-LMSR}
\begin{figure}[htb]
\caption{Tangent line slopes for LS-LMSR (first) and constant product  (second) cost functions}
\label{lslmsrslopefig}
\begin{center}
\includegraphics[width=0.45\textwidth]{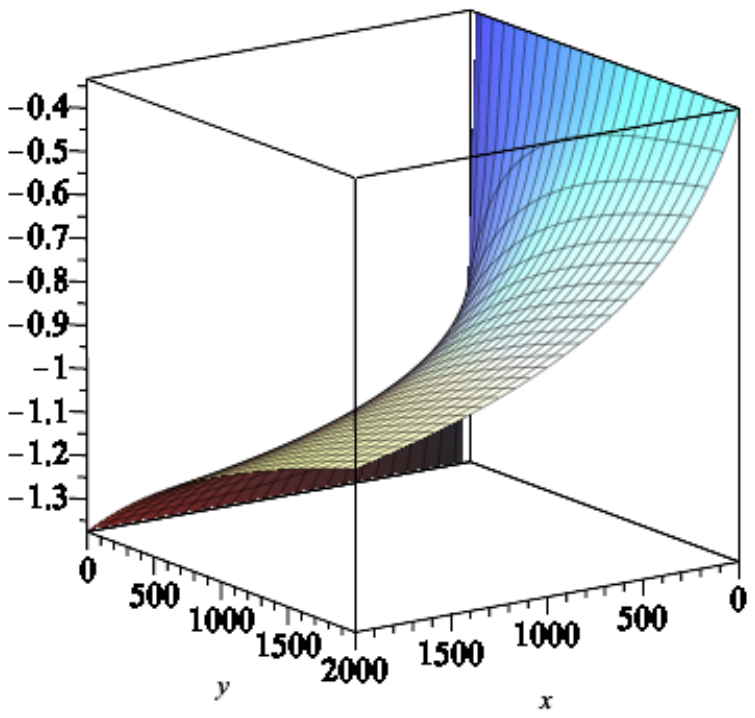} \!\!\!\!\!\!\!\!\!\!\!\!\!\!\!\!\!\!\!\!\!\!\!\!\!\!\!\!\!
\includegraphics[width=0.45\textwidth]{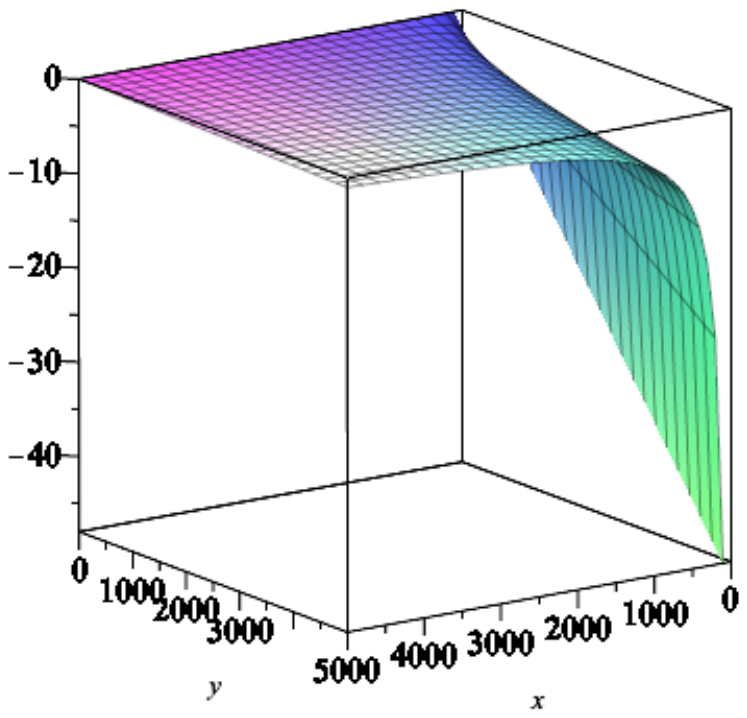} 
\end{center}
\vskip -3cm
\end{figure}

For the LS-LMSR based AMM, the market cost is 
$$C({\bf q}_0)=2000\cdot \ln \left(e^{1000/2000}+e^{1000/2000}\right)=2386.294362.$$
At market state ${\bf q}_0$, the instantaneous prices for a coin of tokens are $P_x({\bf q}_0)=P_y({\bf q}_0)=1.193147181$.
A trader may use 817.07452949 spade suit coins to purchase 1000 USDT coins
with a resulting market state ${\bf q}_1=(0, 1817.07452949)$ and a resulting market cost $C({\bf q}_1)=2386.294362$.
At market state ${\bf q}_1$, the instantaneous prices for a coin of tokens are $P_x({\bf q}_1)=0.8511445298$
and $P_y({\bf q}_1)=1.313261687$. Thus we have ${P_x({\bf q}_1)}/{P_y({\bf q}_1)}=0.6481149479$.
The tangent line slope of the cost function curve indicates the token price fluctuation stability in the
automated market. 
The tangent line slope for 
the LS-LMSR cost function curve at the market state ${\bf q}=(x,y)$ is
$$\frac{\partial y}{\partial x}=-\frac{(x+y)\left(e^{\frac{x}{x+y}}+e^{\frac{y}{x+y}}\right)
\ln\left(e^{\frac{x}{x+y}}+e^{\frac{y}{x+y}}\right)+{y\left(e^{\frac{x}{x+y}}-e^{\frac{y}{x+y}}\right)}}
{(x+y)\left(e^{\frac{x}{x+y}}+e^{\frac{y}{x+y}}\right)
\ln\left(e^{\frac{x}{x+y}}+e^{\frac{y}{x+y}}\right)+{x\left(e^{\frac{y}{x+y}}-e^{\frac{x}{x+y}}\right)}}.$$
For the LS-LMSR AMM with an initial state ${\bf q}_0=(1000, 1000)$,
the tangent line slope (see Figure \ref{lslmsrslopefig}) changes smoothly and 
stays between $-1.542936177$ and $-0.6481149479$.
Thus the token price fluctuation is quite smooth. By the principle of supply and demand, it is expected that 
when the token supply increases, the token price decreases. That is, the cost function 
curve should be convex. However, the cost function curve for LS-LMSR market is concave. 
This can be considered as a disadvantage of LS-LMSR markets for certain DeFi applications. 
Though LS-LMSR does not satisfy the translation invariance property, it is
shown in \cite{othman2013practical} that the sum of prices are bounded by $1+\alpha n\ln n$.
For the two token market with $\alpha=1$, the sum of prices are bounded by $1+2\ln 2=2.386294362 $
and this value is achieved when $x=y$.

As an additional example of LS-LMSR AMMs, a trader may spend 10 USDT 
coins to purchase 10.020996 coins of spade suit token at market state ${\bf q}_0$ or 
spend 500 USDT coins to purchase 559.926783 coins of spade suit from the market state ${\bf q}_0$ 
with a resulting market state $(1500,440.073217)$. Furthermore, in the market state 
$(1500,440.073217)$, the value of one USDT coin is equivalent to the value of 1.260346709 coins of 
spade suit token. 

\subsection{Constant product and constant mean}
For the constant product AMM, the market cost is
$C({\bf q}_0)=1000000$ and the constant product cost function is $x\cdot y=1000000$.
At market state ${\bf q}_0$, the instantaneous token prices are $P_x({\bf q}_0)=P_y({\bf q}_0)=1000$.
Thus we have $\frac{P_x({\bf q})}{P_y({\bf q})}=1$.
A trader may use one USDT coin to buy approximately one coin of spade suit token and vice versa 
at the market state ${\bf q}_0$. However, as market state moves on, the prices 
could change dramatically based on token supply in the market and the pool of a specific coin will never run out.
Specifically, at market state ${\bf q}_0$, a trader may spend 10 USDT coins to purchase 9.900990099
spade suit coins. On the other hand, a user may spend 500 USDT coins to purchase only 333.3333333
coins of spade suit token from the market  state ${\bf q}_0$ with a resulting market state 
${\bf q}_1=(1500,666.6666667)$.
Note that in the example of LS-LMSR market example, at market state ${\bf q}_0$,
a trader can spend 500 USDT coins to purchase 559.926783 coins of spade suit.
Furthermore, in the market state ${\bf q}_1$, one USDT coin could purchase 0.4444444445 coins of 
spade suit token. The tangent line slope of the cost function curve at the market state ${\bf q}=(x,y)$ is
$$\frac{\partial y}{\partial x}=-\frac{P_x({\bf q})}{P_y({\bf q})}=-\frac{y}{x}.$$
That is,  the tangent line slope for the cost function curve (see Figure \ref{lslmsrslopefig})  
can go from $-\infty$ to $0$ and the token price fluctuation could be very sharp. 
Specifically, if the total cost of the initial market ${\bf q}_0$ is ``small'' (compared against attacker's capability),
then a trader/attacker could easily control and manipulate the market price of each coins in the AMM. 
In other words, this kind of market maker may not serve as a reliable price oracle. 
A good aspect of the constant product cost function is that the curve is convex.
Thus when the token supply increases, the token price decreases. 
On the other hand, the sum of prices $P_x({\bf q})+P_y({\bf q})=x+y$ 
in constant product market is unbounded. Thus constant production cost function could not be used in
prediction markets since it leaves a chance for a market maker to derive
unlimited profit from transacting with traders.

For constant mean AMMs, Figure \ref{constantmean3dfig} displays
an instantiated constant mean cost function curve. The curve in Figure \ref{constantmean3dfig} 
is very similar to the curve in Figure \ref{constant3dfig} for the constant product cost function. 
Thus constant mean AMM has similar properties as that for
constant product AMM and we will not go into details.

\subsection{Constant ellipse}
As we have mentioned in the preceding Sections, one may use the upper-right part of the curve 
for a concave cost function or use the lower-left part of the curve for a convex cost function.
In order to conform to the principle of supply and demand, 
we analyze the convex cost functions based on constant ellipse.
Constant ellipse share many similar properties though
they have different characteristics. By adjusting corresponding parameters,
one may obtain different cost function curves with different properties
(e.g., different price fluctuation range, different tangent line slope range, etc). 
The approaches for analyzing these cost function curves are similar. Our following analysis
uses the low-left convex part of the circle $(x-6000)^2 + (y-6000)^2 = 2\times 5000^2$ 
as the constant cost function.

\begin{figure}[htb]
\caption{The tangent line slope for constant ellipse automated market maker}
\label{constantcircleslopefig}
\begin{center}
\includegraphics[width=0.50\textwidth]{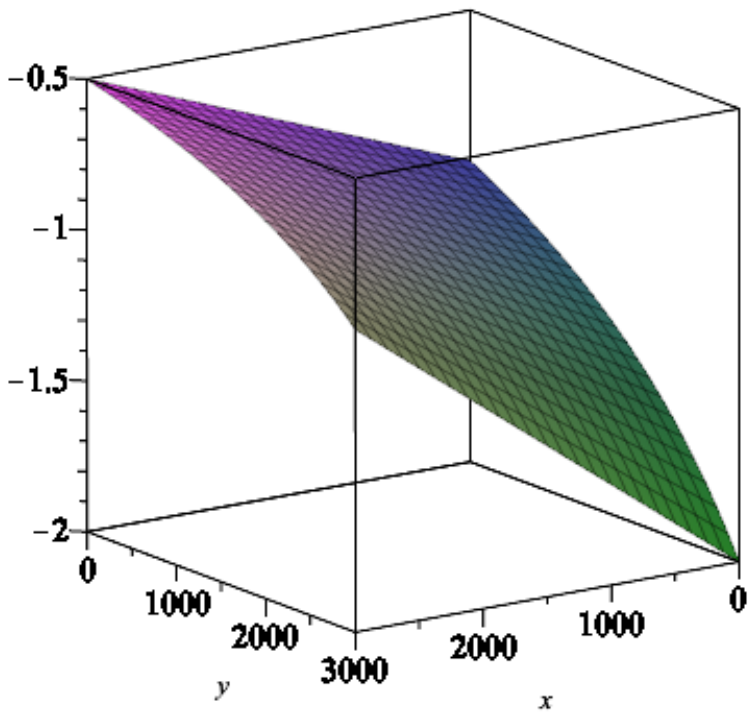}
\end{center}
\vskip -3cm
\end{figure}

For AMMs based on this cost function $C({\bf q})=(x-6000)^2 + (y-6000)^2$, the market cost is 
$C({\bf q}_0)=50000000.$ At market state ${\bf q}_0$, the instantaneous prices for a coin of tokens are 
$P_x({\bf q}_0)=P_y({\bf q}_0)=-10000$.
A trader may use 1258.342613 spade suit coins to purchase 1000 USDT coins
with a resulting market state ${\bf q}_1=(0, 2258.342613)$ and a resulting market cost $C({\bf q}_1)=C({\bf q}_0)$.
At market state ${\bf q}_1$, the instantaneous prices for a coin of tokens are $P_x({\bf q}_1)=12000$
and $P_y({\bf q}_1)=7483.314774$. Thus we have $\frac{P_x({\bf q}_1)}{P_y({\bf q}_1)}=1.603567451$.
The tangent line slope of the cost function curve at the market state ${\bf q}=(x,y)$ is
$$\frac{\partial y}{\partial x}=-\frac{P_x({\bf q})}{P_y({\bf q})}=-\frac{x-6000}{y-6000}.$$
This tangent line slope (see Figure \ref{constantcircleslopefig}) changes smoothly and 
stays in the interval $[-1.603567451, -0.6236095645]$.
Thus the token price fluctuation is quite smooth. Furthermore, this cost function 
has a convex curve which conforms to the principle of supply and demand. That is, 
token price increases when token supply decreases. For constant ellipse cost function market, 
the sum of prices are bounded by $P_x({\bf q})+P_y({\bf q})=2(x+y)-4a$. Similar bounds
hold for constant ellipse cost function market. Thus, when it is used for prediction market, 
there is a limit on the profit that a market maker can derive from transacting with traders.

\begin{figure}[htb]
\caption{Cost functions (bottom up): $(x+y)\ln \left(e^{\frac{x}{x+y}}+e^{\frac{y}{x+y}}\right)=
2000\cdot\ln\left(2e^{1/2}\right)$,
$(x + 6000)^2 + (y + 6000)^2 = 2\times 7000^2$, 
$x+y=2000$, 
$(x-6000)^2 + (y-6000)^2 = 2\times 5000^2$,  and $xy=1000000$}
\label{comparisonfig}
\begin{center}
\includegraphics[width=0.6\textwidth]{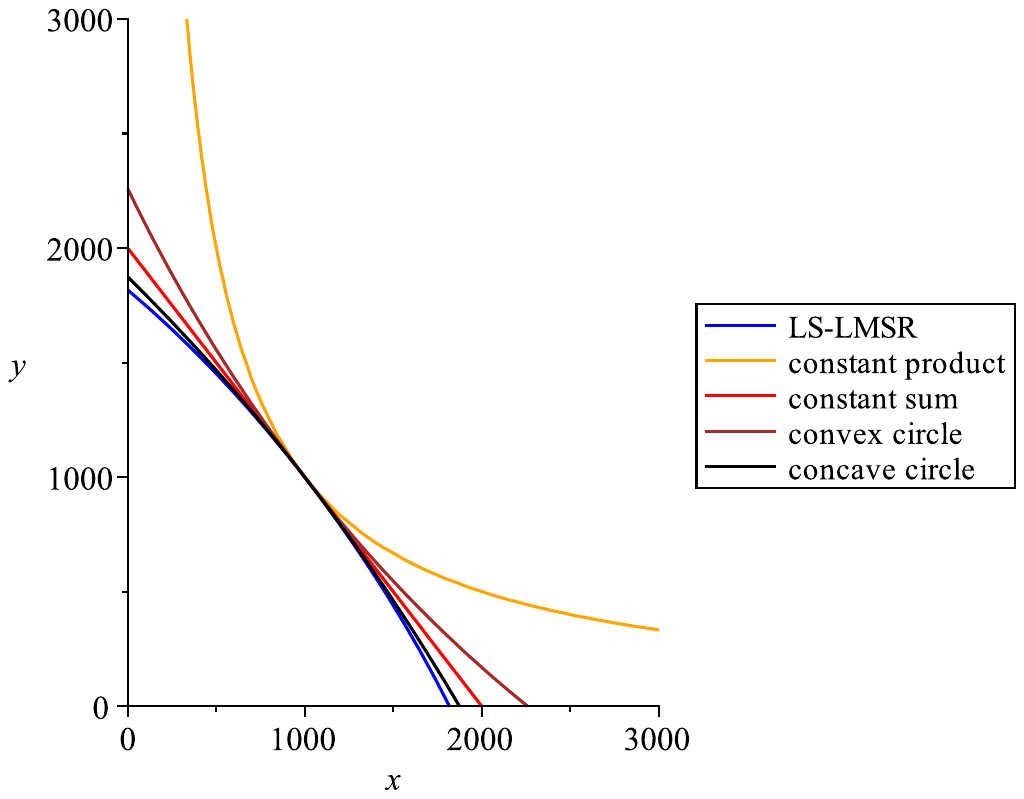} 
\end{center}
\vskip -5cm
\end{figure}

Figure \ref{comparisonfig} compares the cost function curves for different AMMs
that we have discussed. These curves show that constant ellipse cost function
is among the best ones for DeFi applications.

\subsection{Front running attacks based on slippage}
Slippage based front-running attacks can always be launched if the tangent line slope for the cost function curve
is not a constant. The more the tangent line slope fluctuates around the current market state, 
the more profit the front-runner can make. The analysis in preceding sections  show 
that tangent line slopes for LS-LMSR and constant ellipse cost functions fluctuate smoothly
and tangent line slopes for constant product/mean cost functions fluctuate sharply. Thus 
LS-LMSR and constant ellipse cost function automated markets are more robust
against front running attacks.  
In Uniswap V2, when a trader submits a transaction
buying coins of token $A$ with coins of token $B$ (or vice versa), the trader may submit the order at the limit.
But the front runner can always try to profit by letting the trader's order 
be executed at the limit price as shown in the following attacks against Uniswap V2.

\begin{example}
Most front running attacks leverage off-chain bots and on-chain proxy smart contracts to carry out attacks
(see., e.g,  \cite{darkforesteth}). There are some statistics on these front running bots
at Dune Analytics (see, e.g., \cite{collectedbotsDune}). The following two recent proxy smart contracts
take advantage of the large slippage on Uniswap V2.  
\begin{itemize}
\item \href{https://ethervm.io/decompile/0xd59e5b41482ee6283c22e1a6a20756da512ffa97}{\tt 0xd59e5b41482ee6283c22e1a6a20756da512ffa97} received a profit of at least 1,172,436 USD during a 14 days period.
\item \href{https://ethervm.io/decompile/0x000000005736775feb0c8568e7dee77222a26880}{\tt 0x000000005736775feb0c8568e7dee77222a26880} received a profit of 60 ETH during one week. 
The profit was transferred to another address \href{https://etherscan.io/address/0x94dDD5e97de3A659A1b10E2845857eDb01883619}{\tt 0x94dD...}.
\end{itemize}
We analyze attacking steps by the second bot against Uniswap  V2 Pair 
SPA-ETH:
\href{https://v2.info.uniswap.org/pair/0x13444ec1c3ead70ff0cd11a15bfdc385b61b0fc2}{\tt 0x13444...}.
The attacking transactions are included in the block \href{https://etherscan.io/block/12355902}{\tt 12355902}
finalized on May-02-2021 04:43:18 PM.
\begin{enumerate}
\item The attacker saw that \href{https://etherscan.io/address/0x006fa275887292cdc07169f1187b7474e376bb3b}{\tt 0x006fa275887292cdc07169f1187b7474e376bb3b} submitted an order to swap 4.544 ETH for SPA.
\item The attacker's smart contract inserts an order to swap 2.6842 ETH for 385,583 SPA
before the above observed order in transaction hash \href{https://etherscan.io/tx/0x4e2636bb75566ac73150beb9c92c6cbab1342023e907e91d6e93dc0f01635b06/}{\tt 0x4e2636...}
\item \href{https://etherscan.io/address/0x006fa275887292cdc07169f1187b7474e376bb3b}{\tt 0x006fa....}'s 
order is fulfilled at the transaction hash
\href{https://etherscan.io/tx/0x9a17e959255ff7b9ae096c5af0a66992bae5bba055e860b11c32f7114f08e977/}{\tt 0x9a17...}where the user received 613,967 SPA for his 4.544 ETH.
\item The attacker's smart contract inserts an order to swap 385,583 SPA for 2.8778 ETH
after the above observed order  in transaction hash \href{https://etherscan.io/tx/0x34787e325022773d3deb20bdea0b737c0e04aa8f46afa89282c83cc519630388/}{0x34787e...}
\item The attacker's smart contract received 0.1936 ETH for free.
\end{enumerate}
\end{example}

\section{Price amplitude}
\label{lslmsrexsec}
For constant product/mean AMMs, the relative price $\frac{P_1({\bf q})}{P_2({\bf q})}$
of the two tokens ranges from $0$ (not inclusive) to $\infty$. At the time when a tiny portion of one token coin
is equivalent to all coins of the other token in the market maker, no trade is essentially feasible.
Thus the claimed advantage that no one can take out all shares of one token from the constant product/mean market 
seems to have limited value. For a given LS-LMSR (or constant ellipse)  
automated market with an initial state ${\bf q}_0$, the relative price $P_1({\bf q})/P_2({\bf q})$
can take values only from a fixed interval. If the market changes and this relative price interval 
no long reflects the market price of the two tokens, one may need to add tokens to the market 
to adjust this price interval. On the other hand, it may be more efficient to just cancel this automated 
market maker and create a new AMM when this situation happens.

In the following example, we show how to add liquidity to an existing  LS-LMSR AMM
to adjust the relative price range. Assume that the market price for a coin of token $A$ is 
100 times the price for a coin of token $B$ when the AMM is incorporated.
The patron uses 10 coins of token $A$ and 1000 coins of token $B$ to create 
an AMM with the initial state ${\bf q}_0=(1000,1000)$. The total market cost is 
$C({\bf q}_0)=2386.294362$. 
Assume that after some time, the AMM moves to state ${\bf q}_1=(100,1750.618429)$.
At ${\bf q}_1$, we have $P_1({\bf q}_1)/P_2({\bf q}_1)=0.6809820540$
which is close to the lowest possible value 0.6481149479.
In order to adjust the AMM so that it still works 
when the value $P_1/P_2$ in the real world goes below 0.6481149479, the patron can add some 
coins of token $A$ to ${\bf q}_1$ so that the resulting market state is 
${\bf q}_2=(1750.618429,1750.618429)$. To guarantee that 
one coin of token $B$ is equivalent to $\frac{P_2({\bf q}_1)}{100\cdot P_1({\bf q}_1)}
=0.01468467479$ coins of token $A$ in ${\bf q}_2$, we need to have the following mapping 
from outstanding shares in ${\bf q}_2$ to actual token coins (note that this mapping is different from that
for ${\bf q}_0$):
\begin{itemize}
\item Each outstanding share of token $A$ corresponds to 0.01468467479 coin of token $A$.
\item Each outstanding share of token $B$ corresponds to one coin of token $B$.
\end{itemize}
Thus there are $1750.618429\times 0.01468467479=25.70726231$ coins of token $A$
in ${\bf q}_2$. Since there is only one coin of token $A$ in ${\bf q}_1$, the patron 
needs to deposit $24.70726231$ coins of token $A$ to ${\bf q}_1$ to move the AMM
to state ${\bf q}_2$. If the market owner chooses not to deposit
these tokens to the market, the market maker will still run, but there is a chance that the outstanding shares of token
$A$ goes to zero at certain time.

In the above scenario, one may ask whether it is possible for the market maker to automatically 
adjust the market state to ${\bf q}_3=(1750.618429,1750.618429)$ by re-assigning the mapping from shares to coins?
If ${\bf q}_2$ automatically adjusts itself to ${\bf q}_3$ without external liquidity input, 
then a trader may use one share of token $A$ to get one 
share of token $B$ in ${\bf q}_3$. Since we only have one equivalent coin of token $A$ 
but 1750.618429 outstanding shares in ${\bf q}_3$, each outstanding share of 
token $A$ in ${\bf q}_3$ is equivalent to 0.0005712267068 coins of token $A$. That is, the trader 
used 0.0005712267068 coins of token $A$ to get one coin of token $B$ 
(note that each outstanding share of token $B$ corresponds to one coin of token $B$ in ${\bf q}_3$). 
By our analysis in the preceding paragraphs, at ${\bf q}_3$,
one coin of token $B$ has the same market value of 0.01468467479 coins of token $A$.
In other words, the trader used 0.0005712267068 coins of token $A$ to get 
equivalent 0.01468467479 coins of token $A$. Thus it is impossible for the automated market 
to adjust its relative price range without an external liquidity input.

\section{Implementation and performance}
\label{impelsec}
We have implemented the constant ellipse based AMMs  using Solidity smart contracts
and have deployed them over the Ethereum blockchain. The smart contract source codes and Web User Interface 
are available at GitHub.
As an example, we use the ellipse $(x-c)^2 + (y-c)^2=r^2$
to show how to establish a token pair swapping market in this section. Specifically,
we use $c=10^9$ and $r\cdot 10^{14}=16000\cdot 10^{14}$ (that is, $r=16000$) for illustration purpose in this section.

Each token pair market maintains constants $\lambda_0$ and $\lambda_1$  which are determined 
at the birth of the market. Furthermore, each token market also maintains
a non-negative multiplicative scaling variable $\mu$ which is the minimal value 
so that the equation  $(\mu \lambda_0x_0-10^{9})^2 + \left({\mu\lambda_1 y_0}-10^{9}\right)^2\le 16000\cdot 10^{14}$
holds
where $\mu \lambda_0x_0<10^9$ and 
${\mu\lambda_0 y_0}<10^9$. This ensures that we use the lower-left section of the ellipse
for the automated market.  

\subsection{Gas cost and comparison}
We compare the gas cost against Uniswap V2 and Uniswap V3. 
During the implementation, we find out that some of 
the optimization techniques that we used in Coinswap may be used to reduce the gas cost in Uniswap V2. Thus 
we compare the gas cost for Uniswap V2   (column Uni V2)  our optimized version of Uniswap V2 (column Uni V2O),
 Uniswap V3 (column Uni V3),  and our  CoinSwap
in Table \ref{tsfd}. In a summary, our constant ellipse AMM ( CoinSwap) has a gas saving 
from 0.61\% to 46.99\% over Uniswap V2 and has a gas saving from 23.19\% to 184.29\% over Uniswap V3.
It should be noted that Uniswap V3 tried to reduce the slippage in certain categories though still do not have 
the full slippage control as CoinSwap has. The testing script that we have used will be available on the Github.
Some field for Uniswap V3 in Table \ref{tsfd} is empty since we did not find an easy way to test that in the 
Uniswap V3 provided testing scripts.

\begin{table*}
\caption{Gas cost Uniswap V2, V3, and  CoinSwap with liquidity size (40000000,10000000)}
\label{tsfd}
{\small
\begin{center}
\begin{tabular}{ |l|c|c|c|c|c|c|}\hline
function&{\bf UNI V2} & {\bf UNI V2O} & {\bf UNI V3} & {\bf  CoinSwap} & {\bf Saving over UNI V2}& {\bf Saving  over UNI V3}\\ \hline
mint() &141106&132410&308610&109722&28.60\%&184.29\% \\ \hline
swap()&89894&88224&114225&89348&0.61\%&27.84\% \\ \hline
swap()[1st]&101910&100051&&96294&5.83\%& \\ \hline
add $\Omega$&216512&207368&&185442&16.76\%& \\ \hline
remove $\Omega$&98597&97319&82694&67127&46.88\%& 23.19\%\\ \hline
add ETH &223074&213930&&192027&16.14\%& \\ \hline
full removal&123339&122061&&98805&24.83\%& \\ \hline
partial removal&180355&137061&&144283&25.00\%& \\ \hline
\end{tabular}
\end{center}
}
\end{table*}

\section{Conclusion}
\label{conclusionsec}
The analysis in the paper shows that constant ellipse cost functions
have certain advantages for building AMMs in Decentralized Finance (DeFi) applications.
One may argue that constant ellipse cost function based markets have less flexibility after the market
is launched since the price amplitude is fixed. We have mentioned that, though the token price 
could range from 0 to $\infty$ in the constant product cost model, when the price for one token
is close to infinity, any meaningful trade in the market is infeasible. Thus the old market 
needs to be stopped and a new market should be incorporated. Indeed, it is an advantage 
for an AMM to have a fixed price amplitude when it is used as a price oracle for other 
DeFi applications. For the constant product cost market, if the patron incorporates the AMM
by deposing a small amount of liquidity, an attacker with a small budget can manipulate the token price significantly
in the AMM and take profit from other DeFi applications that use this AMM 
as a price oracle. For constant ellipse based AMMs, the patron can use a small amount
of liquidity to set up the automated market and the attacker can only manipulate the token price within
the fixed price amplitude.

\bibliographystyle{plain}

\end{document}